\documentclass[12pt,preprint]{aastex}

\slugcomment{Submitted to ApJ Letters}

\shorttitle{McNeil's Nebula}
\shortauthors{Reipurth and Aspin}

\received{2004 February 18}
\begin{document}


\title{IRAS~05436--0007 and the Emergence of McNeil's Nebula}


\author{Bo Reipurth\altaffilmark{1}}
\affil{Institute for Astronomy, University of Hawaii, 2680 Woodlawn Drive, Honolulu, HI 96822}
\email{reipurth@ifa.hawaii.edu}

\and

\author{Colin Aspin\altaffilmark{2}}
\affil{Gemini Observatory, 670 North A`ohoku Place,\\ University
        Park, Hilo, HI 96720
}
\email{caa@gemini.edu}




\begin{abstract}

We present a study of McNeil's Nebula, a newly appeared reflection
nebula in the L1630 cloud, together with photometry and spectroscopy
of its source. New IR photometry compared to earlier 2MASS data shows
that the star has brightened by about 3 magnitudes in the
near-infrared, changing its location in a J-H/H-K$'$ diagram precisely
along a reddening vector. A Gemini NIRI K-band spectrum shows strong
CO-bandhead emission and Br$\gamma$ is in emission, indicative of
strong accretion. A Gemini GMOS optical spectrum shows only a red,
heavily veiled continuum, with H$\alpha$ strongly in emission and
displaying a pronounced P~Cygni profile, with an absorption trough
reaching velocities up to 600 km~s$^{-1}$. This implies significant
mass loss in a powerful wind. However, no evidence is found for any
shocks, as commonly seen in collimated outflows from young stars.
Apparently the eruption has dispersed a layer of extinction and this,
together with the intrinsic brightening of the IRAS source, has
allowed an earlier outflow cavity to be flooded with light, thus
creating McNeil's Nebula.

\end{abstract}





\section{Introduction}

Wide-field photographic and CCD images have documented the presence of
numerous small compact reflection nebulae in star-forming dark
clouds. Such nebulae may possibly represent a transitional stage in
which a young star goes from being an embedded infrared source to a
visible T Tauri star or Herbig Ae/Be star. This process is widely
believed to involve powerful bipolar outflow activity, as reflected in
the morphology of the nebulae, which are often cometary, with the
partially embedded star illuminating an outflow cavity. When observed
over longer time-scales, such small reflection nebulae are often found
to vary in brightness and occasionally in illumination
pattern. Perhaps the earliest known example is NGC 2261, which
reflects light from the young star R~Mon and was found to vary by
Hubble (1916). More recently, Cohen et al. (1981) demonstrated the
variability of the fan-shaped nebula emanating from PV~Cephei.
At some stage, the outflow activity of a star for the first time
punches a hole in its surroundings, suddenly allowing a beam of light
to sweep across the surroundings. Subsequent eruptions in the source
will again flood the outflow channel with light. The emergence of the
nebula Re~50 in the Lynds 1641 cloud near the embedded source IRAS
05380--0728 was documented by Reipurth \& Bally (1986). Another case
may have been the nebula IC~430 in L1641, which at present is quite
small, but at the end of the 19th century appears to have been a very
large and much brighter object (Pickering 1890; Strom \& Strom 1993).


\section{Observations}

The observations presented below were taken on the "Frederick C. Gillett"
Gemini North Telescope\footnote{Based on observations obtained at the
Gemini Observatory, which is operated by the Association of
Universities for Research in Astronomy, Inc., under a cooperative
agreement with the NSF on behalf of the Gemini partnership: the
National Science Foundation (United States), the Particle Physics and
Astronomy Research Council (United Kingdom), the National Research
Council (Canada), CONICYT (Chile), the Australian Research Council
(Australia), CNPq (Brazil), and CONICET (Argentina).}, on Mauna Kea,
Hawaii on UT dates February 03 (near-IR) and 14 (optical), 2004.  The
near-IR imaging and spectroscopy was acquired with the
facility imager/spectrometer, NIRI, using the f/6 camera giving an
image scale on the Aladdin InSb 1024x1024 detector of 0.116$"$/pixel.
Images in J, H and K$'$ were obtained in total on-source integration
times of 3.6 seconds (0.18 seconds, 10 coadds, two spatial positions)
in each filter.  The data were calibrated using similar observations of
the UKIRT faint standard stars FS113, 119 and 135. The spectroscopic
data were acquired using a 0.5$"$ wide long-slit and K-band grism with a
total on-source integration time of 60 seconds (30 seconds, 1 coadd,
two spatial positions) resulting in a spectrum with resolving power
$\sim$780 covering the wavelength region 2.05-2.45 microns.  Telluric
feature correction was performed using observations of the A0~V star
HIP28056. 
The optical data were acquired using the facility optical multi-object
spectrometer, GMOS-N.  Images in the GMOS Sloan g', r', i' and z'
filters were obtained with exposure times of 60 seconds and 10
seconds, the shorter exposure time being used to give unsaturated
images of the young star.  The GMOS spectrum was a 5~min exposure
obtained using the R831 grating, 0.5$"$ wide long-slit, and a central
wavelength of 5800~\AA\ resulting in a spectrum covering the
wavelength range 4800-6800~\AA\ at a resolving power of R$\sim$4500.
Infrared images were obtained of the eruptive star at the UH 2.2m
telescope through an R-band filter (5 min exposure) on Feb 1, 2004 and
through a narrowband H$_2$ 2.12 $\mu$m filter (20 min exposure) on Feb
14, 2004 UT.

\section{Pre-outburst Observations}

On January 23, 2004, McNeil (2004)\footnote{This reference is listed
as McNeil, Reipurth, \& Meech (2004) in the Astrophysics Data
System. The second and third authors only provided a simple
confirmatory CCD image, and full credit for the important discovery of
the eruption in IRAS~05436--0007 should go solely to
Mr. McNeil. Consequently, we suggest that future reference to the
discovery announcement in this IAU Circular be given as McNeil
(2004).}  discovered the presence of a bright nebula more than an
arcminute in extent south-west of M78 in the L1630 cloud, a nebula
that was not seen on any of the blue or red POSS-I, POSS-II and UKSTU
sky atlas plates taken between 1951 and 1991.  McNeil's Nebula is
clearly cometary-shaped, with a highly reddened, dust-obscured star at
its apex.  Subsequent examination of recent images obtained in the
fall of 2003 suggests that McNeil's Nebula first appeared in late
November 2003 (Brice\~no et al. 2004; McNeil, priv. comm.). However,
Mr. John Welch of Phoenix, Arizona, kindly brought our attention to a
photograph of the M78 region obtained by E. Kreimer on Oct 22, 1966 on
which McNeil's Nebula is seen at virtually the same brightness as at
discovery (Mallas \& Kreimer 1978). This offers the important insight
that the current eruption of the illuminating star is not the first
one. We here summarize what is known about this star prior to the
appearance of the nebula.
The star lies within the error ellipse (45$''$$\times$7$''$ at PA
88$^o$) of the IRAS source 05436--0007, which is detected only at 12
and 25 $\mu$m with fluxes of 0.53 and 1.19 Jy, respectively, with a
flag that indicates an 80\% probability that the fluxes are
variable. The star also coincides with the 2MASS source
05461313--0006048, which is located at $\alpha_{2000}$: 05$^h$46$^m$13.1$^s$,
and $\delta_{2000}$: --00$^o$06$'$05$''$.  The 2MASS photometry is listed in
Table~1.  A faint optical counterpart to this 2MASS source is visible
on an I-band image obtained by Eisl\"offel \& Mundt (1997).
In the sub-millimeter, the source was detected as a compact, isolated
dust continuum source at 350 and 1300 $\mu$m by Lis, Menten, \& Zylka
(1999), their source LMZ~12, a designation we will use for the source
in the following until a variable star name is assigned in the
GCVS. Subsequently, the same source was detected at 850 $\mu$m by
Mitchell et al. (2001) and Johnstone et al. (2001), their source
OriBsmm~55. Lis et al. (1999) suggested a pre-outburst bolometric
luminosity for LMZ~12 of 2.7~L$_\odot$. These sub-millimeter
observations additionally detected extended, diffuse emission about 40
arcsec north of LMZ~12, coinciding with part of McNeil's Nebula.
Limited millimeter observations by Lis et al. (1999) did not reveal
any molecular outflow from the source, and only a small core in
HCO$^+$.

\section{Observational Results}

We present in Fig.~1 a color image of McNeil's Nebula based on g', r',
and i' images taken with GMOS at the Gemini-N telescope in 0.5$''$
seeing.  The nebula has approximate dimensions of 30 $\times$ 60
arcsec, and shows considerable structure, with a very bright patch of
illumination near the source.  Photometry of LMZ~12 at its apex is
given in Table~1. Note that because of the bright nebulosity, the
stellar magnitudes are very sensitive to the aperture-size employed,
we used a small aperture of 0.9$''$ radius.  

The nebula encompasses the Herbig-Haro object HH~22 (Herbig
1974). However, the detection of a collimated jet associated with
HH~22 emanating from a source further to the East suggested to
Eisl\"offel \& Mundt (1997) that there is no connection between this
HH object and LMZ~12. Another HH object, HH~23, is located further to
the North. It consists of two knots on an axis that passes close to
LMZ~12, and Eisl\"offel \& Mundt (1997) suggested that the faint
optical source coincident with LMZ~12 is the driving source. McNeil's
Nebula opens up in the approximate direction towards HH~23, and we
suggest that it represents the illuminated outflow cavity carved at
some time in the past by LMZ~12 into the L1630 cloud. Proper motion
measurements of HH~23 are needed to establish a firmer link with
LMZ~12.

Infrared J, H, K$'$ photometry of the source obtained on Feb 03, 2004
is listed in Table~1. It is evident that the source is much brighter
now than when the 2MASS data were obtained on Oct 07, 1998. At the
time of our observations, LMZ~12 had brightened by $\Delta$J=3.6,
$\Delta$H=3.2, and $\Delta$K$'$=2.9 magnitudes. When plotted in a
J-H/H-K$'$ diagram, it is evident that it displays a substantial
infrared excess (Fig.~2). It is noteworthy that in its current high
state, LMZ~12 shows considerably {\em less} reddening than at the
time of the 2MASS observation in 1998. 

In Fig.~3 we show a contour diagram of LMZ~12 as seen in our K$'$-band
images. It is immediately obvious that the star is surrounded by a
compact reflection nebulosity and that, in particular, this nebula
shows a curved tail characteristic of many stars undergoing
high-accretion events (e.g., Herbig 1977).
In Fig.~4, we show a K-band spectrum of LMZ~12, which displays
a red continuum with strong CO-bandhead emission, and the Br$\gamma$
and Na~I lines are in emission. 
Our optical spectrum of LMZ~12 shows a red continuum with a prominent
H$\alpha$ line in emission, but no other emission or absorption lines,
suggesting the presence of heavy veiling. The H$\alpha$ line, shown in
Fig.~5, has an equivalent width of $-$32~\AA\ and displays a
characteristic P~Cygni profile. The absorption component has an
equivalent width of 5.6~\AA.

\section{Discussion}

The two major classes of eruptive variables among pre-main sequence
stars are the FUors and the EXors (Herbig 1966, 1977, 1989). In the
following we discuss whether LMZ~12 can be linked to one of these two
categories.

FUors are characterized by large-amplitude ($\Delta$V$\sim$5-6 mag.)
brightenings lasting several or many decades, and in the optical they
show F-G type spectra without emission lines. In the K-band region,
FUors display deep CO-bandhead absorption (e.g., Reipurth \& Aspin
1997).  Although LMZ~12 has about the right amplitude, it remains to
be seen how long it stays bright.  Spectroscopically, however, LMZ~12
looks very different from mature FUors both in the optical and in the
infrared. The prominent H$\alpha$ emission and the strong CO emission
that are presently seen do not suggest a classification as a FUor as
we currently understand the phenomenon. On the other hand, Brice\~no
et al. (2004) emphasize that the earliest optical spectrum after
outburst of the FUor V1057 Cyg had some similarities to that of
LMZ~12.  Further photometric and spectroscopic monitoring of the star
is required to settle this.

EXor eruptions, which may occur repeatedly in a given star, have
amplitudes that can be comparable to those of FUors, but they have
durations which are much shorter, from a few months to a few
years. Spectroscopic studies of EXors are limited, because of the
rarity and shorter durations of these eruptions. The few such studies
that have been made (e.g., Lehmann, Reipurth, \& Brandner 1995; Herbig
et al. 2001; Parsamian et al. 2002) all show that EXors in
eruption have at least the lower Balmer lines in emission, together
with emission lines of He~I and other lines characteristic of the most
active T~Tauri stars (Herbig 1962). A high-resolution spectrum of the
H$\alpha$ line of SVS~13 in outburst (Eisl\"offel et al. 1991) shows a
P~Cygni profile rather similar to the one seen in Fig.~5.  However, we
do not see any other emission lines than H$\alpha$ in the spectral
range observed (4800-6800~\AA) in LMZ~12.  Infrared spectra of EXors
are even more limited, one case is the eruption of SVS~13, for which a
K-band spectrum shows the CO bandheads strongly in emission as well as
Br$\gamma$ in emission (Carr \& Tokunaga 1992).
However, a similar spectrum of EX~Lup shows Br$\gamma$ in emission,
but the CO bandheads and Na~I in absorption (Herbig et al. 2001). It
thus does not seem that EXors have a unique infrared spectral
signature.
Based on a comparison with these somewhat limited observations of
EXors, we conclude that LMZ~12 does have a certain resemblance to
EXors.

The spectroscopic data give us some insight into the physical
processes behind the observed eruption.  The detection of Br$\gamma$
in emission testifies to a region of hot gas. Br$\gamma$ emission has
been shown to correlate tightly with accretion luminosity, so its
presence indicates that the eruption is linked to an episode of
accretion (Najita et al. 1996; Muzerolle et al. 1998).  The pronounced
P~Cygni profile seen at H$\alpha$ in the optical spectrum is likely to
be formed in a strong wind that has sufficient optical depth to
produce the deep blueshifted absorption trough (e.g., Muzerolle et
al. 2001). Only few T~Tauri stars show such well-developed P~Cygni
profiles at H$\alpha$. The well-defined blue edge of the absorption
trough indicates wind velocities of up to 600 km~s$^{-1}$, even more
than the extreme wind that emanates from FU~Orionis (e.g., Herbig et
al. 2003).

It is noteworthy that we see no spectral features indicative of
shocks: the infrared spectrum shows no evidence for H$_2$ emission,
and the optical spectrum does not display the [SII] 6717/6731 lines
characteristic of Herbig-Haro jets. Our H$_2$ 2.122 $\mu$m image of
LMZ~12 shows no presence of any extended jet flow. It thus appears
that LMZ~12, in common with most FUors and EXors, has powerful mass
loss, but apparently not in the well-collimated fashion that enables
shocked HH jets. 


One final insight into the LMZ~12 EXor event comes from the changes of
the infrared colors from before to after the eruption. The pre- and
post-outburst colors show that LMZ~12 has moved precisely along a
reddening vector, indicating that the star has brightened partly
because its visual extinction diminished by about 4.5
magnitudes. Using the reddening curve of Rieke \& Lebofsky (1985), we
find that this corresponds to A$_J$=1.26, A$_H$=0.81, and
A$_K$=0.50. Since we know that the infrared colors changed by
$\Delta$J=3.64, $\Delta$H=3.16, and $\Delta$K=2.87, we see that the
intrinsic brightening of the star corrected for the change in
extinction is 2.4 magnitudes in each of the J, H, and K filters.
Through a combination of this intrinsic brightening and the clearing
away of an obscuring layer, the star has been able to illuminate its
outflow channel, thus creating McNeil's Nebula.


\acknowledgments

We are grateful to Nuria Calvet for helpful comments, Mike Connelley
for taking an infrared H$_2$ image, George Herbig for illuminating
discussions, Inger Jorgensen and Kathy Roth for a GMOS spectrum, Jay
McNeil for alerting us to the outburst, Karen Meech and Yan Fernandez
for a CCD image, Peter Michaud and Kirk Pu'uohau-Pummill for preparing
the color-image, Jean-Ren\'e Roy for approving all the Gemini 8m ToO
observations, Brian Skiff for information about both recent and
pre-discovery images, John Welch for information about the 1966
outburst, and an anonymous referee for helpful comments.

\clearpage

\begin{deluxetable}{lrrrrrrr}
\tablecaption{Photometry of IRAS~05436--0007\label{tbl-1}}
\tablewidth{0pt}
\tablehead{
\colhead{Date} & \colhead{g'} & \colhead{r'} & \colhead{i'} & 
\colhead{J} & \colhead{H} & \colhead{K$'$} & \colhead{Telescope}
}
\startdata
Oct 7, 1998  & & & &  14.74$\pm$.03 & 12.16$\pm$.03 & 10.27$\pm$.02 & 2MASS \\
Feb 03, 2004$^a$   & & & &  11.1$\pm$.1   &  9.0$\pm$.1   &  7.4$\pm$.1 & Gemini \\ 
Feb 14, 2004$^b$ & 22.8 & 17.4 & 15.6 &  & & & Gemini\\
\enddata


\tablenotetext{a}{We list large and conservative error estimates for
the Gemini IR photometry, since this object is very bright for an 8m
telescope, requiring very brief integrations.}  
\tablenotetext{b}{GMOS photometry is obtained in an aperture with
radius 0.9$''$ in order to avoid bright nearby nebulosity.  It is
quoted to one decimal place only since we use standard zeropoints from
the Gemini GMOS web site.}

\end{deluxetable}


\clearpage

\begin{figure}
\includegraphics[angle=0,scale=1.8]{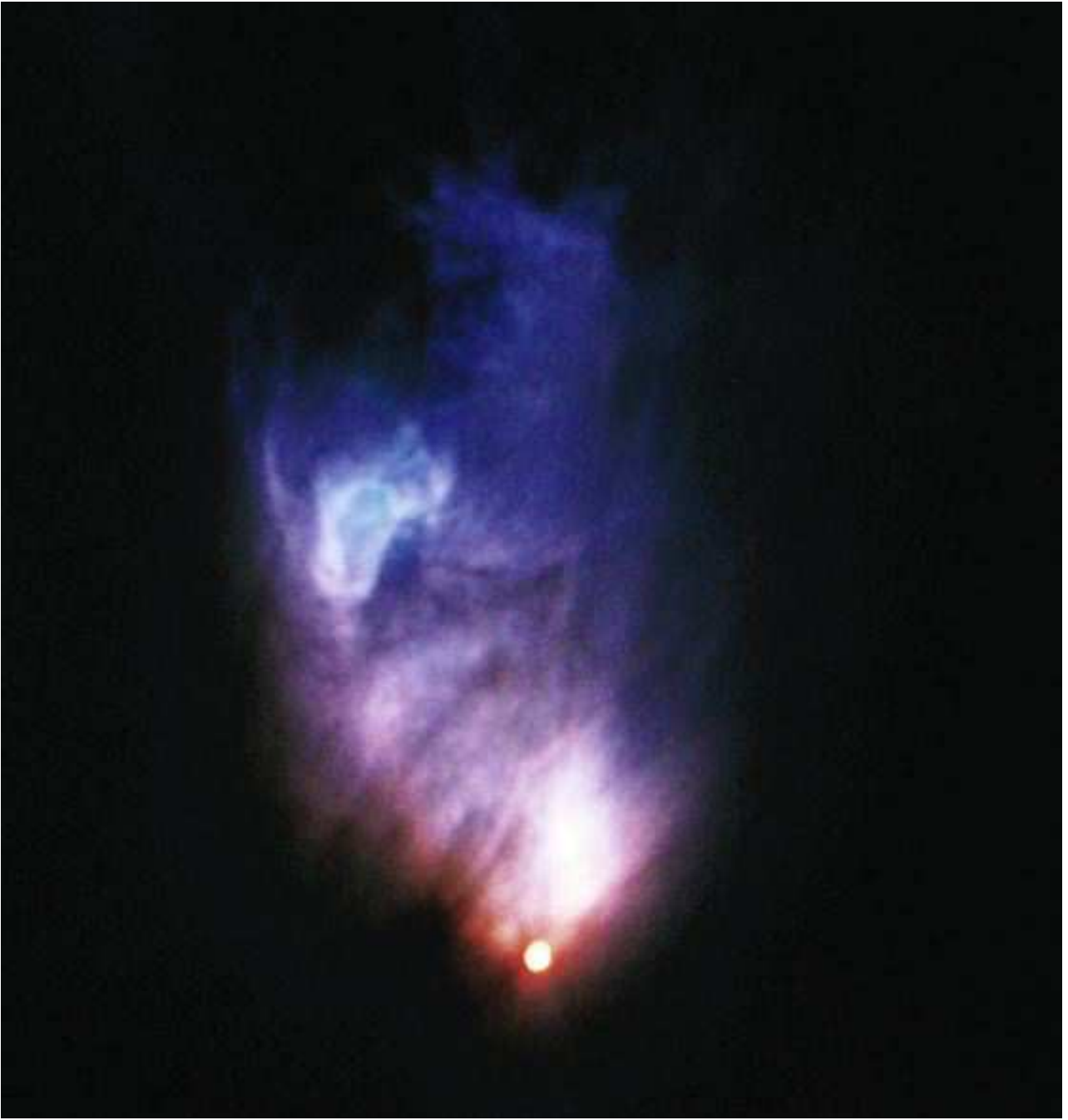}
\caption{A color-image of McNeil's Nebula obtained by combining
broadband g', r', and i' images obtained with GMOS at the Gemini-N 8m
telescope. LMZ~12 is at the bottom of the nebula, and HH~22 is the
bluish curved object on the northeastern edge of McNeil's Nebula. The
height of the figure is about 80 arcsec. North is up and East is
left.
\label{fig1  }}
\end{figure}

\begin{figure}
\includegraphics[angle=0,scale=.75]{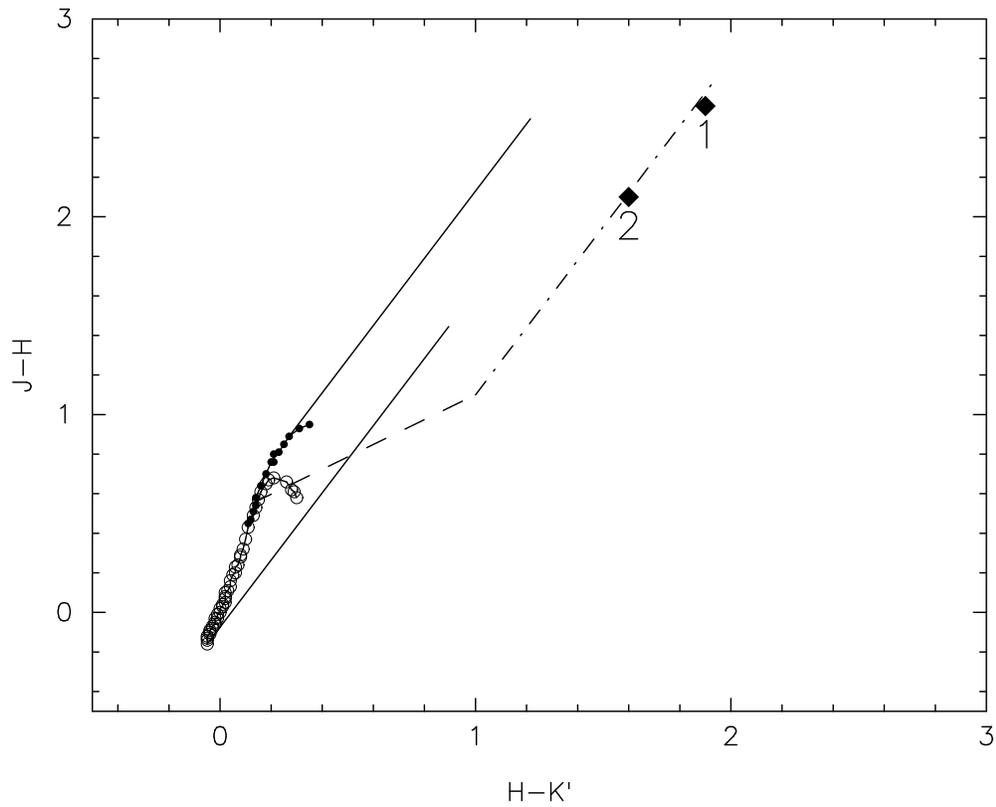}
\caption{A J-H/H-K$'$ diagram showing the location of LMZ~12 as observed
with 2MASS [1] on Oct 7, 1998 and with Gemini-N [2] on Feb 3, 2004.
The dashed line is the T Tauri locus, and the solid straight lines and
dot-dashed line are reddening vectors of A$_V$=15 mag.
\label{fig2}}
\end{figure}

\begin{figure}
\includegraphics[angle=0,scale=.75]{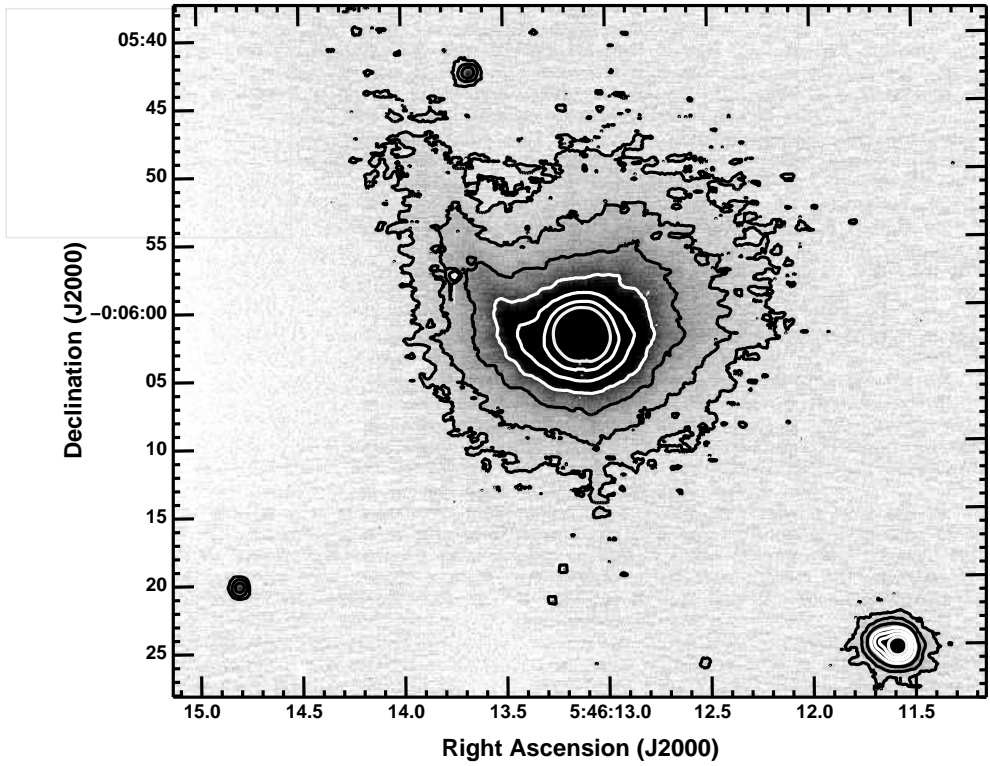}
\caption{A compact nebula is seen around the illuminating star of
McNeil's Nebula in this K$'$-band image obtained with NIRI at the
Gemini-N 8m telescope.
\label{fig3}}
\end{figure}

\begin{figure}
\includegraphics[angle=0,scale=.75]{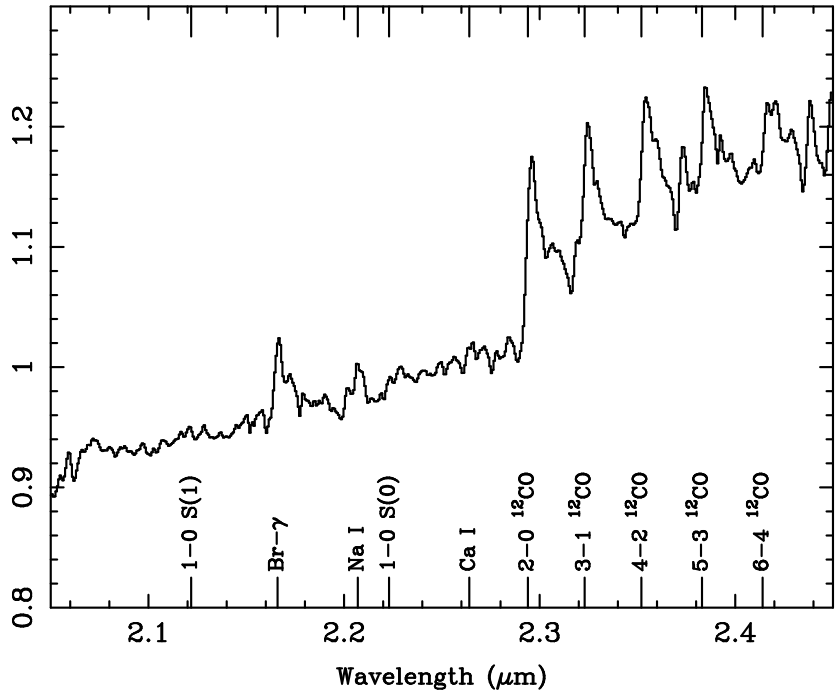}
\caption{A K-band spectrum of LMZ~12 obtained with NIRI at the Gemini-N
8m telescope. \label{fig4} }
\end{figure}


\begin{figure}
\includegraphics[angle=0,scale=.75]{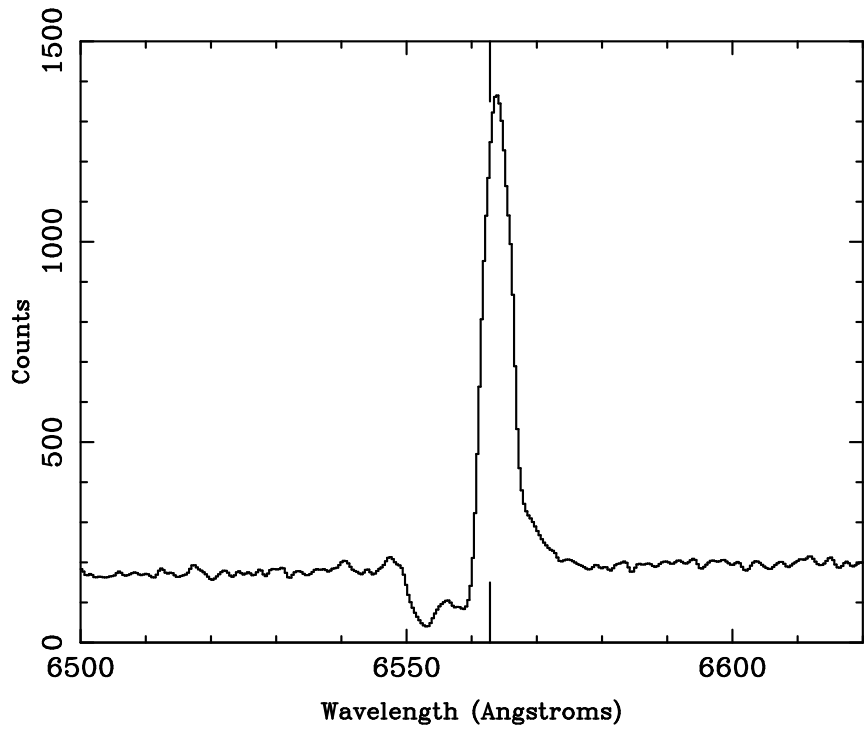}
\caption{The H$\alpha$ line in LMZ~12 shows a pronounced P~Cygni
profile. The rest velocity of the star is indicated. \label{fig5} }
\end{figure}


\begin{thebibliography}{}


\bibitem[]{} Brice\~no, C. et al. 2004, ApJ Letters, in prep.

\bibitem[]{600} Carr, J.S., \& Tokunaga, A.T. 1992, ApJ, 393, L67

\bibitem[]{602} Cohen, M., Kuhi, L.V., Harlan, E.A., Spinrad, H. 1981,
ApJ, 245, 920

\bibitem[]{605} Eisl\"offel, J., \& Mundt, R. 1997, AJ, 114, 280

\bibitem[]{607} Eisl\"offel, J. et al. 1991, ApJ, 383, L19




\bibitem[]{618} Herbig, G.H. 1962, Adv. Astr. Astrophys., 1, 47

\bibitem[]{620} Herbig, G.H. 1966, { Vistas in Astronomy}, 8, 109

\bibitem[]{622} Herbig, G.H. 1974,  Lick Obs. Bull.  No. 658

\bibitem[]{624} Herbig, G.H. 1977, ApJ, 217, 693 

\bibitem[]{626} Herbig, G.H. 1989, in ESO Workshop on {\em Low Mass Star
Formation and Pre-Main Sequence Objects}, ed. B. Reipurth, p. 233

\bibitem[]{629} Herbig, G.H., Aspin, C., Gilmore, A.C., Imhoff, C.L.,
Jones, A.F. 2001, PASP, 113, 1547

\bibitem[]{632} Herbig, G.H., Petrov, P.P., \& Duemmler, R. 2003, ApJ, 595, 384

\bibitem[]{634} Hubble, E.P. 1916, ApJ, 44, 190

\bibitem[]{636} Johnstone, D., Fich, M., Mitchell, G.F.,
Moriarty-Schieven, G. 2001, ApJ, 559, 307


\bibitem[]{643} Lehmann, T., Reipurth, B., Brandner, W. 1995, A\&A, 300, L9

\bibitem[]{645} Lis, D.C., Menten, K.M., \& Zylka, R. 1999, ApJ, 527, 856 

\bibitem[]{647} Mallas, J.H., \& Kreimer, E. 1978, {\em The Messier Album}, Sky Publishing Corporation

\bibitem[]{649} McNeil, J.W. 2004, IAUC 8284

\bibitem[]{651} Mitchell, G.F., Johnstone, D., Moriarty-Schieven, G., Fich, M., Tothill, N.F.H. 2001, ApJ, 556, 215

\bibitem[]{653} Muzerolle, J., Calvet, N., \& Hartmann, L. 1998, AJ, 116, 2965

\bibitem[]{655} Muzerolle, J., Calvet, N., \& Hartmann, L. 2001, ApJ, 550, 944

\bibitem[]{657} Najita, J., Carr, J.S., \& Tokunaga, A.T. 1996, ApJ, 456, 292

\bibitem[]{659} Parsamian, E.S., Mujica, R., \& Corral, L. 2002,
Astrophysics, 45, 393

\bibitem[]{662} Pickering, E.C. 1890, Ann. Harvard College Obs., 18, 113

\bibitem[]{666} Reipurth, B., \& Aspin, C. 1997, AJ, 114, 2700

\bibitem[]{670} Reipurth, B., \& Bally, J. 1986, Nature, 320, 336


\bibitem[]{675} Rieke, G.H., \& Lebofsky, M.J. 1985, ApJ, 288, 618

\bibitem[]{677} Strom, K.M., \& Strom, S.E. 1993, ApJ, 412, L63

\end{thebibliography}
\end{document}